\documentstyle[12pt,psfig,aaspp4]{article}

\received{25 August 1997}

\lefthead{Pedersen et al.}
\righthead{Diverse optical signatures ...}

\begin{document}

\title{
Evidence for Diverse Optical Emission from Gamma-Ray Burst Sources }

\author{H.Pedersen} 
\affil{{Copenhagen University Observatory,
Niels Bohr Institute for Astrophysics, Physics, and Geophysics, 
Juliane Maries Vej 30, DK 2100 Copenhagen, Denmark  } }

\author{{A.O.Jaunsen, and T.Grav}}
\affil{ {Institute for Theoretical Astrophysics, Oslo, Norway} } 

\author{{R.\O stensen, and M.I.Andersen}}
\affil{ {Nordic Optical Telescope, La Palma, Spain} } 

\author{{M.Wold, H.Kristen, A.Broeils, M.N\"aslund, and C.Fransson}}

\affil{ {Stockholm Observatory, Sweden} } 

\author{{M.Lacy}}
\affil{{Nuclear and Astrophysics Laboratory, Oxford, UK} } 

\author{{A.J.Castro-Tirado, and J.Gorosabel}}
\affil{{Laboratorio de Astrof\'{\i}sica Espacial y
F\'{\i}sica Fundamental, INTA, Madrid, Spain} } 

\author{{J. M. Rodriguez Espinosa, and A.M.Perez}}
\affil{{Instituto de Astrof\'{\i}sica de Canarias, Tenerife, Spain} } 

\author{{C.Wolf, and R.Fockenbrock}}
\affil{{Max-Planck Institute f\" ur Astronomie, Heidelberg, Germany}  } 

\author{{J.Hjorth}}
\affil{{Nordic Institute 
for Theoretical Physics, Copenhagen, Denmark} }

\author{{P.Muhli and P.Hakala}}
\affil{{University of Helsinki, Finland} } 

\author{{L.Piro, M.Feroci, and E.Costa}}
\affil{{Istituto di Astrofisica Spaziale, CNR, Frascati, Italy} } 

\author{{L.Nicastro, and E.Palazzi}}
\affil{{Istituto Tecnologie e Studio Radiazioni Extraterrestri, CNR, Bologna, Italy} }

\author{{F.Frontera, and L.Monaldi}}
\affil{{Universita di Ferrara, Italy} } 

\and

\author{{J.Heise}}
\affil{{Space Research Organization of the Netherlands, Utrecht, The Netherlands} }



\def\figenv#1#2#3#4#5{
\begin{figure}[#1]
\vspace{#2}
\begin{center}
\begin{minipage}{12cm}
\caption[#5]{#3 \label{#4.fig}}
\end{minipage}
\end{center}
\normalsize
\end{figure}
}

\def\tabenv#1#2#3#4#5#6{
\begin{table}[#1] 
\begin{center}
\caption[#6]{#2 \label{#5.tab} }
\begin{tabular}{#3}
\vspace{-2mm} \\
\hline 
\vspace{-2mm} \\
#4
\vspace{-2mm} \\
\hline \hline 
\vspace{-2mm} \\
\end{tabular}
\end{center}
\end{table}
}


\newpage

\begin{abstract}

Optical Transients from gamma-ray burst sources, 
in addition to offering a distance determination,
convey important information on
the physics of the emission mechanism, and perhaps
also about the underlying energy source.
As the gamma-ray phenomenon is extremely diverse,
with time scales spanning several orders of magnitude,
some diversity in optical counterpart signatures appears
 plausible.

We have studied the Optical Transient, which 
accompanied the gamma-ray burst of May 8, 1997 (GRB 970508).
Observations conducted at the 2.5-m Nordic Optical Telescope (NOT)
and the 2.2-m telescope at the German-Spanish Calar Alto observatory (CAHA)
cover the time interval starting 3 hours 5 minutes to
96 days after the high energy event. This brackets all
other published observations, including 
radio. When analyzed in conjunction
with optical data from other observatories,
evidence emerges for a composite light curve.
The first interval, from 3 to 8 hours after the event
was characterized by a constant, or slowly declining brightness.
At a later moment the brightness started
increasing rapidly, and reached a
maximum approximately 40 hours after the GRB. From that moment  
the GRB brightness decayed approximately as a power-law
of index -1.21.
The last observation, after 96 days, m$_R$ = 24.28 $\pm$ 0.10, 
is brighter than the extrapolated power-law,
and hints that a constant component, 
m$_R$ = 25.50 $\pm$ 0.40 is present. 
The OT is unresolved (FWHM 0.83\arcsec ) 
at the faintest magnitude level. 

The brightness of the optical transient, its duration and the general
shape of the light curve sets this source apart 
from the single other optical transient known, 
that of the February 28, 1997 event. 

\end{abstract}


\newpage
\section{    Beppo-SAX  }

The launch of the Italian/Dutch 
BeppoSAX satellite, on April 30, 1996, facilitated
the swift calculation of accurate GRB positions, and thereby
the search for rapidly decaying multiwavelength counterparts.
Five such opportunities have been reported:
GRB 960720, 970111, 970228, 970402, and 970508. The localizations
and numerous follow-up studies were published in IAU Circulars
and elsewhere. 

Before GRB 970508, the best studied BeppoSAX event 
was GRB 970228 (\cite{ref01}). This led to the first detection of an 
associated X-ray transient (\cite{ref02}),
an optical transient (\cite{ref03}), 
and an infrared source (\cite{ref04}; \cite{ref05}), all 
at consistent positions.
The earliest optical detection (\cite{ref06}; \cite{ref07}) was made
 15.4 hours
after the high energy event (T+15.4$^h$).
Thereafter the optical source declined monotonically in 
brightness, and was a factor 40 fainter in  the R-band 
by T+6$^d$. 

The detection of an underlying extended source, ostensibly
a galaxy (\cite{ref03}), led to the conclusion
that GRB 970228 is extragalactic. However, the claim by \cite{ref08}
that the OT shows proper motion 
(which is disputed by \cite{ref09}\footnote{An HST observation,
conducted September 4, and reported by A.Fruchter at the
Fourth Huntsville Symposium on 
Gamma-Ray Bursts, further argues
against proper motion}) 
argues for a galactic origin. 
 The possibility that the extended
source is variable (\cite{ref100}) would 
further complicate interpretation, 
but also this result has been questioned (\cite{ref11}). 

Upper limits to the optical emission have been derived 
for two other events, GRB 970111 (\cite{ref12}) and 970402 
(\cite{ref13}; \cite{ref14}).  
For the first event, observations started 19 hours after the 
GRB and continued until one month later. Any fading was $<$ 0.2 mag for 
objects with m$_B$ $<$ 21, m$_R$ $<$ 20.8 and $<$ 0.5 mag for those down 
to m$_B$ = 23, m$_R$ = 22.6. Thereby the source differs significantly from 
GRB970228.
The second event, which was followed by an X-ray transient (\cite{ref15}), 
is located at low galactic latitude,
b$^I$$^I$ = 9$^o$, for which reason the deepest optical 
data do not constrain the emission beyond that 
known from GRB 970228.

\section{    GRB 970508    }

The rather weak gamma-ray burst 
GRB 970508 was detected (\cite{ref16})  by BeppoSAX 
on May 08, 1997 UTC 21 h 41 min 48 s; it lasted 15 s.
It was recorded also by the BATSE detectors on-board 
Compton Gamma Ray Observatory (\cite{ref17}). 
At the higher sensitivity of that instrument, 
the duration was 35 s, including a single pulse of
3.6 s FWHM.

Within three hours, the source was located
to a precision of 10\arcmin . 
This led to the successive 
detection of a transient (decaying) X-ray source 
(Piro et al. 1997a;c),
a variable optical source (\cite{ref20}), and a 
radio source (\cite{ref21}; \cite{ref22}), all at consistent positions. 
The radio source was initially undetected, at T+3.7$^h$ and T+22.5$^h$,
and seen rising from T+5.0$^d$ to T+6.2$^d$. VLBA measurements 
(\cite{ref22})
show that it is unresolved to 0.0003\arcsec \ (at 8.4 GHz).
The closest galaxies detected by Palomar and HST observations
are several arcseconds away (\cite{ref23}; \cite{ref30}), and
they have a large probability of being present due to chance.

At two epochs of spectroscopic measurements 
the redshift z=0.835 was deduced, first from absorption lines,
then also from emission of [O II] 372.8 nm
(Metzger et al. 1997b;c).
This constitutes the first distance determination for a Gamma-Ray Burst.
Assuming a redshift z=1, \cite{ref26} estimated
the total energy output as 10$^{52}$ ergs.

\section{ NOT data               }

Observations at the NOT (+ALFOSC, HiRAC)
began May 9 UTC 00 h 47 min 18 s, i.e. 3 h, 5 min after the GRB 
(see the observing log, Table 1). 
The ALFOSC detector was a thinned Loral 2K CCD, with a spectral range
of 320-1050 nm. Multiplied with the atmospheric- and instrumental 
transmission functions
we estimate that the sensitivity function has a maximum
between 600 and 700 nm, 
and a FWHM range from 450 to 800 nm 
(http://www.not.iac.es/instruments/alfosc/ccd$\_$sens.html). 
When used without a filter, this represents a wide R band,
hereafter referred to as R\arcmin \ .  
Observations in V (recalibrated since first published, \cite{ref28})
and I are listed for reference only; they are too
few to merit separate discussion.

The last observations were done August 13 - 15, 1997, i.e. 95 to 97
days after the event. Figure 1 represents the digital sum of all 13 exposures
done in that time interval. The image stack has been corrected for cosmic 
rays, using a skewness criterion.
On this image, the FWHM of two field stars is 0.83\arcsec \, that 
of the OT 0.79\arcsec \ (i.e.\ consistent within errors).
There is an extremely faint intensity maximum, conceivably due to
statistical noise, $\sim$1\arcsec \ W of the OT.
The 13 measurements of the OT brightness
relative to 
object A have r.m.s.\ = 0.30 mag, fixing m$_R$ = 24.28 $\pm$ 0.10.

\section{      CAHA data         }

Observations with the 2.2-m CAHA telescope (+CAFOS)
began May 9, UTC 01 h 48 min (Castro-Tirado et al. 1997); 
they include unfiltered as well as R(Kron-Cousins) data.

\section{    Discussion   }

Figure 2 shows the combined red and unfiltered photometry,
including data obtained by 
\cite{ref291},
 \cite{ref29}, 
Djorgovski et al. (1997) and
Pian et al.\ (1997)
 and
from IAU Circulars 6654 through 6676. 
Most, possibly all, R photometry refers to the Kron-Cousins
system (cw 659 nm, FWHM 157 nm, \cite{ref77}).
For the unfiltered HST/STIS observation (Pian et al. 1997;
Sahu et al. 1997)  
we have used the simulated R magnitude.
Gunn r photometry differs mainly by the narrower transmission peak
of this system (cw 654 nm, FWHM 89 nm). 
There is little indication that
the merging of data from different color-systems influences 
the scatter of the data.
If anything, the NOT R\arcmin \ data 
should be shifted by $\Delta$m $\sim$ --0.15 mag,
to bring them into better agreement with other data.
However, we have no other justification for such transformation.

The rise to maximum is composed by an initial
period of constancy or modest fall in brightness, 
documented between T+3$^h$ and T+8$^h$,
and a fast rise, seen between epochs T+22$^h$ and T+30$^h$.
The significance of the early decay hinges on  
the unknown color-indices of the OT as function of time. 
We do, however, find it difficult
to reconcile the early part of the light curve with
a {\it rise} in luminosity.

Subsequent to the maximum near T+40$^h$ the brightness declined
approximately as a power law; this has already been
discussed by Pian et al.\ (1997). 
However, the last observation, at T+96$^d$, m$_R$ =  24.28 $\pm$ 0.10,
is significantly brighter than indicated by the trend of other data. 
This indicates that a source of constant luminosity  
contributes to the total light.
We have solved for this contribution, using a weighted least 
squares fit, and find m$_R$ = 25.50 $\pm$0.40, where
the error includes an estimated uncertainty of the fit.
The OT itself would then decline as m$_R$ = 18.95 + 3.03 log(T),
corresponding to index -1.21. 
Alternatively, the power-law,
albeit well founded theoretically (e.g. \cite{ref33}; 
Chiang \& Dermer 1997), does not apply over 
such a  long range. 
We do not detect a visibly extended image profile, which 
argues in favor of the departure from the power-law decay 
being intrinsic to the OT rather than due to the appearance 
of an underlying host galaxy, although an underlying compact 
dwarf galaxy may not have been resolved by our observations. 
The upper limit known from HST data (\cite{ref09}) 
is 25.5, and thus consistent with NOT data.

It is noteworthy that by T+3$^h$, 
the OT was $\sim$3 magnitudes brighter than the 
level to which it had declined by T+96$^d$.
In the most plausible scenario, this early 
brightness would be reached as a consequence of the 
high energy event, rather than being its progenitor.

Eruptive astrophysical phenomena showing two light-curve 
maxima include supernovae.  However, the involved time scales 
are much longer, $\sim$78 days, for the radioactivity 
induced {\it exponential} decay (see e.g. \cite{ref31}).

Some gamma-ray burst models (e.g.\ \cite{ref32})
 predict an early 
phase of multiwavelength radiation, simultaneous to the high 
energy event, others a delayed maximum 
(e.g. \cite{ref33}; \cite{ref34}), and still others, both 
(e.g. \cite{ref35}).
Several models involve both internal and external shocks.
\cite{ref36} and Wijers (personal communication) 
hint at the possibility that a fast jet impacting
on slower, yet relativistic material around it
may produce curves with different time scales,
of which the slower one never makes gamma rays.
The initial decay of the present OT light curve -if real-
may therefore be more closely related to the central engine of the GRB
than the later peak.

\cite{ref292} have fitted a model of beamed, superrelativistic 
outflow to published R photometry of this source. 
Optical observers viewing the outflow 
from an angle $\theta$ $>$ 1/$\Gamma_0$
(where $\Gamma_0$ is the initial bulk Lorenz factor) 
would see an almost constant luminosity starting at
the moment of the GRB, followed by a rise
to maximum when the Lorenz factor has decreased to $\sim$ 1/$\theta$.
The early data from NOT and CAHA conform well to this  model,
but we note that the rise to maximum was 
faster than described by theory, and the decay slower.
To explain this, we propose that the interstellar medium near the
GRB is inhomogeneous. Like all models involving beaming, a high
rate of events is required, compared to the isotropic emission
models. 

In Table 2 we compare some of the characteristics
of GRB 970228 and 970508. While the $\gamma$- and X-ray
flux of the latter event was weaker by a factor $\sim$5,
its optical signature was stronger (factor $\sim$ 2) 
and it reached maximum with greater delay (factor $>$2).
As both sources are located off the galactic plane
(at b$^{II}$ = -17.9$^o$ and +26.7$^o$, respectively), 
we can therefore conclude that Optical Transients
from gamma-ray burst sources show considerable
diversity. 

\acknowledgments

We are grateful to the director 
of the Nordic Optical Telescope for his continued
support for this program.
HP acknowledges support from the Danish Space Research Institute
and Copenhagen University Observatory.
J.Kemenen is thanked for advance information on Konkoly observations,
A.Kopylov for standard stars data.
The German--Spanish Astronomical Center, Calar Alto, Spain, is
operated jointly by the Max-Planck-Institut fur Astronomie (MPIA),
Heidelberg, and the  Comisi\'on Nacional de Astronom\'{\i}a, Madrid.
The Elba workshop on Gamma-Ray Bursts 
provided a timely occasion for the discussion of this event.

\begin{figure}
\psfig{figure=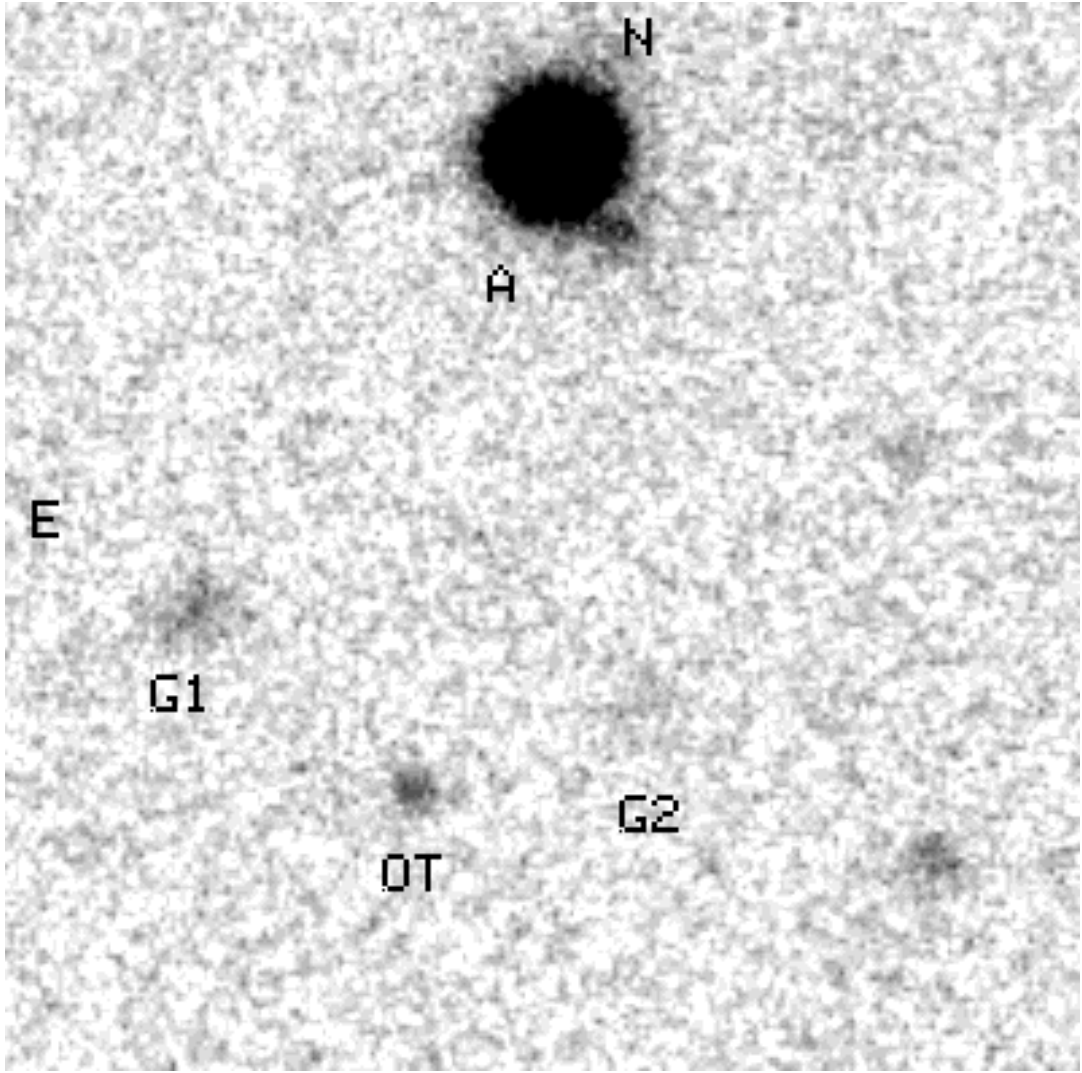,width=20cm}
\caption[]{
The digital sum of all exposures, August 13 to 15, 1997.
The field is 22\arcsec \ by 22\arcsec . }
\end{figure}

\begin{figure}
\psfig{figure=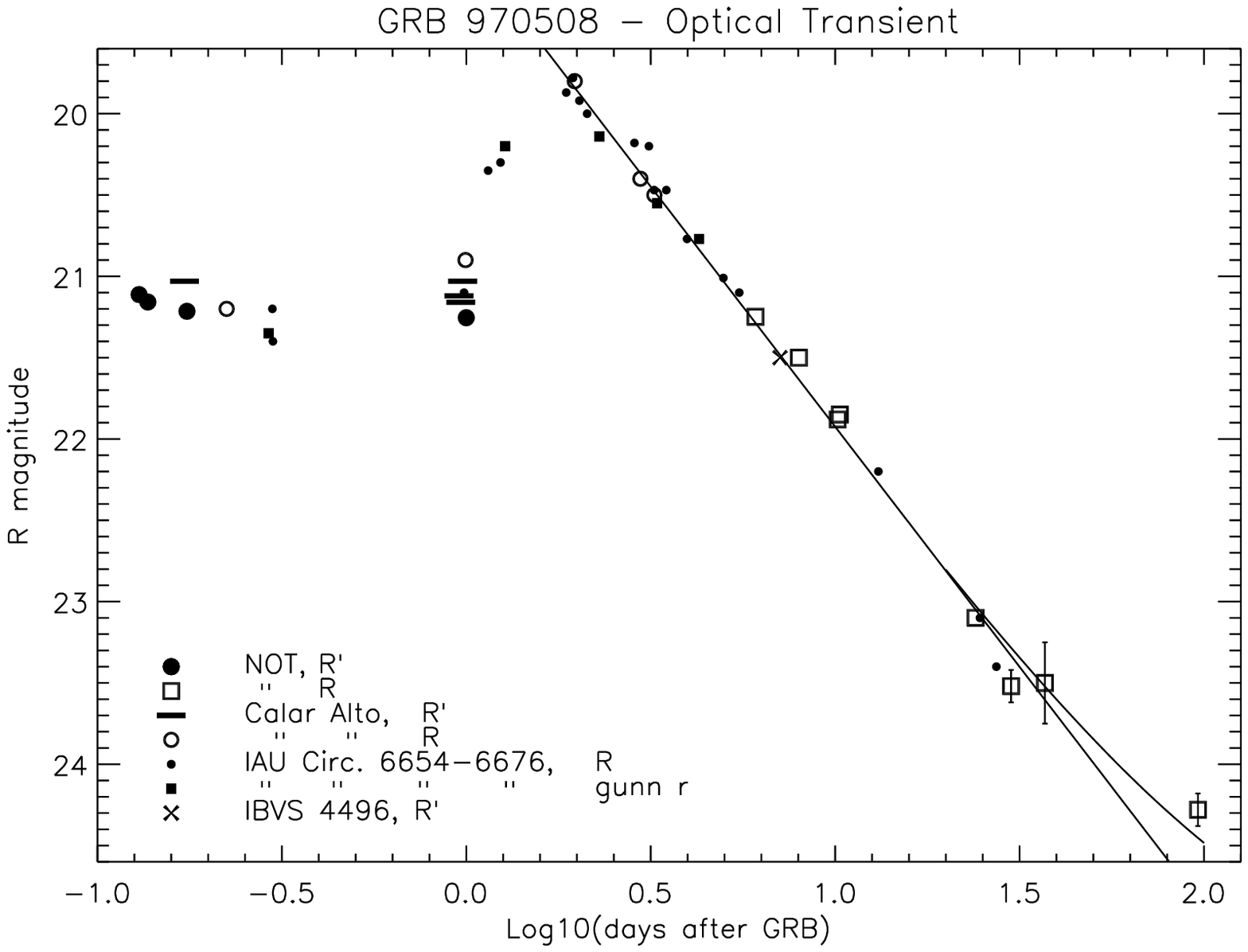,width=15cm}
\caption[]{
The R-band magnitude of the Optical Transient of 
GRB 970508 as function of time after the event. 
For clarity, error bars are only applied to the latest data.

Following a period of constancy (or modest decline), a
peak of optical emission was reached some 40 hours after the
GRB. During the subsequent 35-day interval, the emission decayed 
approximately as a power law. 
The possible contribution of a constant source, m$_R$ = 25.50,
is shown by the curved line.}
\end{figure}

\newpage
\noindent
Table 1:
\vskip 5mm
\begin{tabular}{llrrl}
\hline
\hline
Start, UTC &  Filter & Expos. [s] & magnitude & error    \\ 
\hline
NOT: \\
May 09 00 h 47 min 18 s  & R\arcmin \     &   300   & 21.11  &  0.08    \\
May 09 00 h 54 min 19 s  & R\arcmin \     &   600   & 21.16  &   0.05   \\
May 09 01 h 48 min 50 s  & R\arcmin \     &   300   & 21.21  &   0.06   \\
May 09 21 h 37 min 05 s  & R\arcmin \     &   300   & 21.25  &   0.06   \\
May 10 21 h 37 min 36 s  & V     &   300   & 20.00  &   0.03 \\
May 10 22 h 13 min 50 s  & V     &   600   & 19.96  &   0.03 \\
May 10 22 h 37 min 03 s  & I     &   600   & 19.16  &   0.03 \\
May 11 22 h 15 min 20 s  & V     &   300   & 20.45  &   0.03 \\
May 11 22 h 22 min 00 s  & I     &   300   & 19.48  &   0.03 \\
May 14 23 h 29 min 22 s  & R     &   600   & 21.25  &   0.05\\
May 16 21 h 12 min 30 s  & R     & 900, 180& 21.51  &   0.10 \\
May 19 01 h 13 min 46 s  & R     &  1800   & 21.88  &   0.25\\
May 19 04 h 26 min 23 s  & R     &  2700   & 21.92  &   0.10 \\
June 1 21 h 53 min 15 s  & R     & 1200    & 23.10  &   0.07  \\
June 7 21 h 06 min 06 s  & R     &1200, 600 & 23.52  &   0.10 \\
June 14 22 h 13 min 33 s  & R   & 3 x 1200 & 23.50  &   0.25 \\
Aug 09 04 h 03 min 07 s  &  R    & 3 x 600  & $>$23.70  & (**)    \\
Aug 13.17 -- Aug 15.19  &  R    & 13 x 600  & 24.28  &   0.10 \\
\hline
CAHA (*): \\
May 09 20 h 36 min 28 s & R$\arcmin \ $ &   600   & 21.12  &   0.05  \\
May 09 20 h 52 min 25 s & R$\arcmin \ $ &   600   & 21.16  &   0.05  \\
May 09 21 h 09 min 21 s & R$\arcmin \ $ &   600   & 21.03  &   0.05  \\
\hline

\end{tabular}

\vskip 3mm
\noindent
* One earlier R\arcmin \ measurement published by \cite{ref29}

\noindent
** Seeing poor, OT not clearly detected.

\newpage
\noindent
Table 2:
\vskip 5mm

\noindent
\begin{tabular}{lll}
\hline
\hline
          & GRB 970228   & GRB 970508 \\
\hline
$\gamma$-ray duration (SAX)   &   80 s (ref.\ a)     &       15 s (ref.\ b)    \\
$\gamma$-ray fluence, 40-700 keV (*) & 1.1 (ref.\ c)  &    0.18 $\pm$ 0.03 (ref.\ d)    \\ 
X-ray flux  0.5-10 keV (**) &  4.0 $\pm$ 0.6 (ref.\ e)  & 0.7 $\pm$ 0.08 (ref.\ f)    \\
X-ray flux,  2-10 keV (**) &  2.8 $\pm$ 0.4 (ref.\ e)  & 0.63 $\pm$ 0.06 (ref.\ f)    \\
                                      & (@ 8 hour delay) & (@ 5.7 h delay)  \\

Max  brightness, m$_R$ & 20.7 (ref.\ g)      &  19.8          \\
Optical peak  & $\leq$ 18 hours  (ref.\ g)             & 40 hours        \\
\hline

\end{tabular}

\vskip 03mm
\noindent
a: Costa et al. (1997a); 
b: Costa et al. (1997c); 
c: \cite{ref37}; 
d: Piro et al. (1997c); 
e: Costa et al. (1997b);
f: Piro et al. (1997a); 
g: Galama et al. (1997).

\noindent
*  unit 10$^{-5}$ ergs cm$^{-2}$  

\noindent
** unit 10$^{-12}$ ergs cm$^{-2}$ s$^{-1}$

\end{document}